\documentclass[12pt,a4paper,fleqn]{article}
\usepackage[utf8x]{inputenc}
\usepackage[english]{babel}
\usepackage{slashed}
\usepackage[T1]{fontenc}
\usepackage{cite}
\usepackage{float}
\usepackage[font=small,labelfont=md]{caption}
\usepackage{hyperref}
\usepackage{amsmath,amsthm,amssymb,latexsym}
 \usepackage[english]{babel}
\usepackage[dvips]{graphicx}
\usepackage{amsfonts}
\usepackage{textcomp}
\usepackage{subfig}
\usepackage{pstricks}
\usepackage{authblk}
\usepackage[normalem]{ulem}
\usepackage{color}
\usepackage[title,titletoc,toc]{appendix}
\def\ns{\hspace*{-0.5cm}}
\title{The  top quark  right coupling in the tbW-vertex}

\author[1]{Gabriel A. Gonz\'alez-Sprinberg\thanks{gabrielg@fisica.edu.uy}}
\author[2]{Jordi Vidal\thanks{vidal@uv.es}}

\affil[1]{Instituto de F\'\i sica, Facultad de Ciencias, Universidad de la
Rep\'ublica, Igu\'a 4225, Montevideo 11600, Uruguay.}
\affil[2]{Departament de F\'\i sica Te\`orica, Universitat de Val\`encia, and
Instituto de F\'\i sica Corpuscular (IFIC), Centro Mixto Universitat de Val\`encia-CSIC, E-46100
Burjassot, Val\`encia, Spain.}

\begin{document}

\maketitle
\vspace {-11cm}
\hfill  FTUV - 15 - 1006.2801
\vspace {10cm}

\begin{abstract}

The most general parametrization of the $tbW$ vertex includes a right coupling $V_R$ that is zero at tree level in the standard model. This quantity may be measured  at the Large Hadron Collider where the physics of the top decay is currently investigated.  This coupling is present in new physics models at tree level and/or through radiative corrections, so its measurement can be  sensitive to non standard physics. In this paper we compute the leading electroweak and QCD contributions to the top $V_R$ coupling in the standard model. This value is the starting point in order to separate  the standard model effects and, then, search for new physics. We also propose  observables that can be addressed at the LHC in order to measure this coupling. These observables are defined in such a way that they do not receive tree level contributions from the standard model  and are directly proportional to the right coupling. Bounds on new physics models can be obtained through the measurements of these observables.

\end{abstract}

\section{Introduction}

Top quark physics is now a high statistics physics, mainly due to the huge amounts of data coming from the Large Hadron Collider (LHC) run I and, now, run II \cite{Schilling:2012dx,Hawkings:2015ega}. 
It is strongly believed that, due to its very high mass, the top quark will be a window to new physics \cite{Bernreuther:2008ju}. This can be easily understood in the effective lagrangian approach, where the new physics contributions can be parametrized in a series expansion in terms of the parameter $m_t/\Lambda$, where $m_t$ is the top quarks mass and $\Lambda$ is the new  physics scale. Besides, most of the top quark properties and couplings are known with a precision far below than the other quarks and than the other standard model (SM) particles. 
It is the only quark that weakly decays before hadronization but, up to now, only one decay mode, $t \rightarrow b W^+$, is known. For instance, many extensions of the SM predict new decay modes that can be accessible at LHC. The top quark was detected for the first time at TEVATRON\cite{Abe:1995hr,Abachi:1995iq}, where  many of its  physical properties where measured and some bound on the anomalous $Wtb$ couplings were set \cite{Deterre:2012vn, Abazov:2011pm,Abazov:2008sz}. Nowadays, top physics is intensively investigated  in theoretical research \cite{Bernreuther:2008ju} and at the LHC \cite{Schilling:2012dx,Hawkings:2015ega,Bernardo:2014vha}, as the reader can  verified in the ATLAS and CMS web pages. In particular, the measurements of the different helicity components of the $W$ in the top decay allows to study the $tbW$ Lorentz vertex structure\cite{delAguila:2002nf}. These studies were extended in recent years  \cite{AguilarSaavedra:2010nx,Drobnak:2010ej,Rindani:2011pk,Cao:2015doa,Prasath:2014mfa}.  There, the longitudinal and transverse helicities of the $W$ coming from the top decay were investigated and they show that a precise  determination of the  Lorentz form factors of the vertex  can be done with a suitable choice of observables. The most general parametrization of the on-shell vertex needs four couplings, but in the SM three of them are zero at tree level while only the usual left, coupling $V_L$ is not zero and with a value close to one \cite{Agashe:2014kda}. This is not the case in extended models where, in addition, some of these couplings can also be sensitive to new CP-violation mechanisms. 

The right top coupling is largely unknown and deserves a careful study. 
The other two couplings, usually called tensorial couplings, were investigated
at the LHC\cite{MorenoLlacer:2014tca} and will not be considered here. The  predictions for the SM,   for the two-Higgs-doublet model (2HDM) and other extended models  where recently considered in refs. \cite{GonzalezSprinberg:2011kx,Duarte:2013zfa,Bernreuther:2008us}. In this paper we  first compute the right coupling $V_R$ in the
SM at leading order, and define appropriate observables in order to have direct access to it. The former one loop calculation is
 needed in order to disentangle SM and new physics effects in the observables. Next, we will introduce a set of observables that allows to perform a precise search of the  $V_R$  coupling. We obtain a combination of observables directly proportional to it
 in such a way that they are not dominated by the leading tree level SM
contribution $V_L$. These observables can be an important tool in order to measure new physics contributions to  $V_R$.

This paper is organized as follows. In the next section we introduce a precise definition of the right  coupling  and compute the first order QCD and electroweak (EW) contributions.  In section 3 we present and discuss  a set of  observables that can be measured with LHC data, both in the polarization matrix and in  the spin correlations. Finally, we discuss the results and present our conclusions in section 4.

\section{Right top \texorpdfstring{$tbW$}{tbW} coupling in the SM}\label{gesA2HDM}

Considering the most general Lorentz structure for on-shell particles, the $\mathcal{M}_{tbW}$ amplitude for the
$t(p)\rightarrow b(p') W^{+}(q)$ decay can be written in the following way:
\begin{eqnarray}
\mathcal{M}_{tbW}&=& - \frac{e}{\sin\theta_{w}\sqrt{2}}\epsilon^{\mu*} \, \times \nonumber \\
&&\hspace*{-0.5cm} \overline{u}_{b}(p')\left[\gamma_{\mu}(V_{L} P_{L}+V_{R} P_{R})+ 
\frac{i \sigma_{\mu\nu}q^{\nu}}{M_{W}}( g^{}_{L} P_{L}+ g^{}_{R} P_{R})\right]u_{t}(p),\label{gesdef}
\end{eqnarray}
where the outgoing $W^{+}$ momentum, mass and polarization vector are $q=p-p'$,  $M_{W}$ and $\epsilon^{\mu}$, respectively.
The form factors $V_{L}$ and $V_{R}$ are the left and right couplings, respectively,    while $ g^{}_{L}$ and $ g^{}_{R}$ are  the so called tensorial  or anomalous couplings.

The expression (\ref{gesdef}) is the most general model independent parametrization for the $tbW^{+}$ vertex.  Within the SM, the $W$ couples only to left particles then, at tree level, all the couplings are zero except for $V_L$ that is given by the Kobayashi-Maskawa matix element $V_L=V_{tb} \simeq 1$ \cite{Agashe:2014kda}. At one loop all of these couplings receive contributions from the SM. For example, the real and imaginary parts of the  $g_L$ and $g_R$ couplings were calculated in the SM \cite{GonzalezSprinberg:2011kx} and in a general alligned-2HDM \cite{Duarte:2013zfa}.

\begin{figure}[ht]
\begin{center}
\includegraphics[width=8cm]{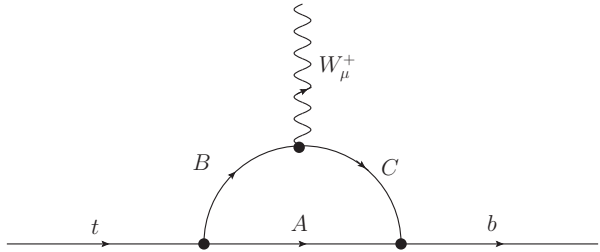}
\end{center}
\caption{One-loop contributions to the $V_R$ coupling in the $t\rightarrow b W^+$ decay.}
\label{figura1}
\end{figure}

 The one loop electroweak and QCD contributions to the $V_R$ coupling can be calculated just by considering the  vertex corrections shown in figure \ref{figura1} and extracting from them the Lorentz structure corresponding to the  $V_R$ coupling.
 Note that this Lorentz structure is not present in the SM lagrangian so, at one loop, there is no need for any counter-term for this contribution. For this reason, the SM one-loop contribution to the $V_R$ coupling is finite. This is similar to what happens for the  top $g_{L,R}$ couplings\cite{GonzalezSprinberg:2011kx,Duarte:2013zfa}.
We will denote each diagram by the label $ABC$ according with the particles running in the loop. In addition to the usual notation for the SM particles we use the symbols $w_0$ and $w$ for the neutral and charge would-be Goldstone bosons, respectively, and $H$ for the SM Higgs. There are 18 diagrams that have to be considered and the one loop $V_R$ value one gets from them is finite, without the need of renormalization as already stated. In particular, the one-loop contributions from diagrams $tw_0w$, $tHw$, $bww_0$, $bwH$, $w_0tb$ and $Htb$ are ultraviolet (UV) divergent. However, when summing them up by pairs ({\it i.e.} $tw_0w + tHw$, $bww_0 + bwH$ and  $w_0tb + Htb$ ), the result is finite. The  electroweak contributions of all the  diagrams are given in the appendix \ref{app:AppendixA} in terms of parametric integrals. For the UV divergent diagrams  we present the sum of the  two diagrams that cancels the divergence, as can be seen  in  eqs. (\ref{div1}), (\ref{div2}) and (\ref{Htb}). There, the first (second) term, in each of these expressions, corresponds to the UV safe sector of the first (second) diagram, while the third one corresponds to the sum of the UV divergent part of the two  diagrams that results in a finite contribution, as it should be.  

All the contributions can be written as:
\begin{equation}
V_R^{ABC} = \alpha\; V_{tb}\; r_b\, I^{ABC},
\label{vr}\end{equation}
where $r_b=m_b/m_t$ and $I^{ABC}$ is an integral shown in appendix \ref{app:AppendixA} for all the diagrams that contribute to $V_R$.
As expected,  all the contribution are proportional to the bottom mass through $r_b$.

When one of the particles circulating in the loop is a photon the integral can be performed analytically. Then, the contribution can be written as
\begin{equation}
V_R^{ABC}= \frac{\alpha }{8 \pi}V_{tb}\, Q_A\, I_0^{ABC},
\label{m0}\end{equation}
with $Q_A$ being the charge of the $A$-quark circulating in the loop in units of $|e|$ (for the $\gamma tb$ diagram, $Q_A=Q_t\cdot Q_b$). 
The $I_0^{ABC}$ analytical expressions, as well as their limits for $r_b \rightarrow 0$, are shown in appendix \ref{app:AppendixB}. All these expressions were used as a check of our computations.

The leading QCD contribution can be easily obtained from eq. (\ref{gtb}) just by substituting the couplings of the photon by the gluon ones, so we get:
\begin{equation}
V_R^{gtb}= -\frac{\alpha_s}{8\pi}\, C_F\, V_{tb} I_0^{\gamma tb},
\end{equation}
with $I^{\gamma tb}$ given in eq. (\ref{Igtb}) and (\ref{Igtb1}), and $C_F=4/3 $ is the color factor.

With the set of values of ref.\cite{Agashe:2014kda}, the numerical values for the contribution to $V_R$ of each diagram and the complete one-loop SM value are given in table \ref{tabla1}, for $V_{tb}=1$. 
\begin{table}[hbt]
{\centering
\caption{Contribution to $V_R$ from the different diagrams \label{tabla1}}
\begin{center}
\begin{tabular}{||c|c|c||}
\hline\hline
Diagram& \multicolumn{2}{|c||}{Contribution to $V_R$}  \\ \hline
$tZW$& \multicolumn{2}{|c||}{$2.01\times 10^{-5}$}\\ \hline
$t\gamma W$& \multicolumn{2}{|c||}{$-1.10\times 10^{-5}$}\\ \hline
$tHW$&\multicolumn{2}{|c||}{ $0$}\\ \hline
$tw_0w$&$\big[(-3.05\times 10^{-5})$&\\ 
+&$+(0.64\times 10^{-5})$&$=-1.55\times 10^{-5}$\\ 
$tHw$&$+(0.86\times 10^{-5})\big]$&\\ \hline
$tZw$&\multicolumn{2}{|c||}{$0.10\times 10^{-5}$}\\ \hline
$t\gamma w$&\multicolumn{2}{|c||}{$0.69\times 10^{-5}$}\\ \hline
$bWZ$&\multicolumn{2}{|c||}{$(1.12+8.24\, i)\times 10^{-5}$}\\ \hline
$bW\gamma$&\multicolumn{2}{|c||}{$(8.34-4.25\, i)\times 10^{-5}$}\\ \hline
$bWH$&\multicolumn{2}{|c||}{$0$}\\ \hline
$bww_0$&$\big[(-0.72+3.73\, i\times 10^{-5})$&\\ 
+&$+(-0.10-2.99\, i\times 10^{-5})$&$=(1.01-0.35\, i)\times 10^{-5}$\\ 
$bwH$& $+(1.83-1.09\, i\times 10^{-5})\big]$&\\ \hline
$bwZ$&\multicolumn{2}{|c||}{ $ (0.00+0.31\, i)\times 10^{-5}$}\\ \hline
$bw\gamma$&\multicolumn{2}{|c||}{$(-4.47+2.29\, i)\times 10^{-5}$}\\ \hline
$Ztb$ & \multicolumn{2}{|c||}{$-2.30\times 10^{-5}$}\\ \hline
$\gamma tb$&\multicolumn{2}{|c||}{ $-2.78\times 10^{-5}$}\\  \hline
$w_0tb$& $\big[(-0.24\times 10^{-5})$&\\ 
+& $+(0.20\times 10^{-5})$&$=-1.03\times 10^{-5}$\\ 
$Htb$ & $+(-0.99\times 10^{-5})\big]$&\\
\hline\hline
$\Sigma(EW)$&\multicolumn{2}{|c||}{ $(0.06+6.23\, i)\times 10^{-5}$}\\ 
\hline \hline
$gtb(QCD)$& \multicolumn{2}{|c||}{$2.68\times 10^{-3}$}\\
\hline\hline
$(QCD+EW)$&\multicolumn{2}{|c||}{$(2.68+0.06\, i)\times 10^{-3}$}\\
\hline\hline
\end{tabular}
\end{center}
}
\end{table}

In this table the contribution from the UV divergent diagrams are summed up to get a finite result; the first (second) quantity between brackets corresponds to the finite contribution of the first (second) diagram, and the third one (which is logarithmic) is the finite sum of the two UV
divergent parts. 

As can be seen in table \ref{tabla1}, even though the contribution of most of the  diagrams is of the order of $10^{-5}$ for the real part of the $V_R$ coupling, the total EW contribution is, at the
end,  two orders of magnitude smaller due to the accidental cancellations among the diagrams. The $QCD$ contribution is real and  four orders of magnitude bigger than the real EW one so that the real part of the $V_R$ coupling is dominated by the former. The imaginary part, instead,  remains of order $10^{-5}$, and it is purely EW.

\section{Observables}\label{Observables}

In general, the LHC observables considered  in the literature  are not very sensitive to the right
coupling $V_R$. This is due to the fact that this coupling comes from a lagrangian term that  has the same parity and chirality properties than the  leading coupling $V_L$  so that the observables receive contributions from both terms.  These observables are the angular asymmetries in the $W$ rest
frame\cite{Lampe:1995xb,delAguila:2002nf,AguilarSaavedra:2006fy, AguilarSaavedra:2010nx}, angular asymmetries in the top  rest frame \cite{AguilarSaavedra:2006fy,Grzadkowski:1999iq,Godbole:2006tq,AguilarSaavedra:2010nx} and spin correlations
\cite{Stelzer:1995gc,Mahlon:1995zn,AguilarSaavedra:2006fy,AguilarSaavedra:2010nx}. Our strategy will be to define observables directly proportional to $V_R$ considering the dependencies on the coupling terms. Similar
ideas were widely applied when investigating tau physics dipole moments \cite{Bernabeu:1993er,Bernabeu:1994wh}. For top decays, one way to suppress the $V_L$ contribution is to define
observables where only right polarized quarks contribute, but this  polarization is not accessible to the present facilities and experiments. Given the results shown in the previous section, from now on we always assume that the
imaginary part of the $V_R$ coupling is negligible. 

\subsection{Observables in the \texorpdfstring{$W$}{W} rest frame}

Top properties were studied in previous works  by means of the
observables that we just mentioned. One of the first possibilities are the angular asymmetries for the  $t\rightarrow
W^+\, b$ decay, with the $W^+$ decaying leptonicaly.  The
normalized charged lepton angular distribution in the W rest frame  can be written as:
\begin{equation}
\frac{1}{\Gamma}\frac{d\Gamma}{d\cos\theta_l}=\frac{3}{8}(1+\cos\theta_l)^2F_++\frac{3}{8}(1-\cos\theta_l)^2F_- +\frac{3}{4}\sin^2\theta_l\, F_0,
\end{equation}
where $F_0, F_\pm$ are the normalized partial widths of the top decay into the  W helicity states, and   $\theta_l$  is the angle between the charged lepton momentum in the $W$ rest frame and the $W$ momentum in the $t$ rest frame. Then, the  asymmetries  are defined, in terms of a new parameter $z$, as follows \cite{Lampe:1995xb,delAguila:2002nf}:
\begin{equation}
A_z=\frac{N(\cos\theta_l >z)-N(\cos\theta_l <z)}{N(\cos\theta_l >z)+N(\cos\theta_l <z)}.\label{eq:asim1}
\end{equation}
 The $z$ parameter allows to separate the  helicity fractions $F_0, F_\pm$.  The value   $z=0$ gives the usual forward-backwards asymmetry while $z=\pm(2^{2/3}-1)$ defines the $A_\mp$ asymmetries, respectively. The first one, $A_0$ depends  only on the  $F_\pm$ helicity fractions while  $A_{\pm}$  depends  on  both $F_0$ and $F_\pm$.

\begin{figure}[ht]
\centering
\begin{minipage}[t]{0.44\textwidth}
\centering
\includegraphics[width=\textwidth]{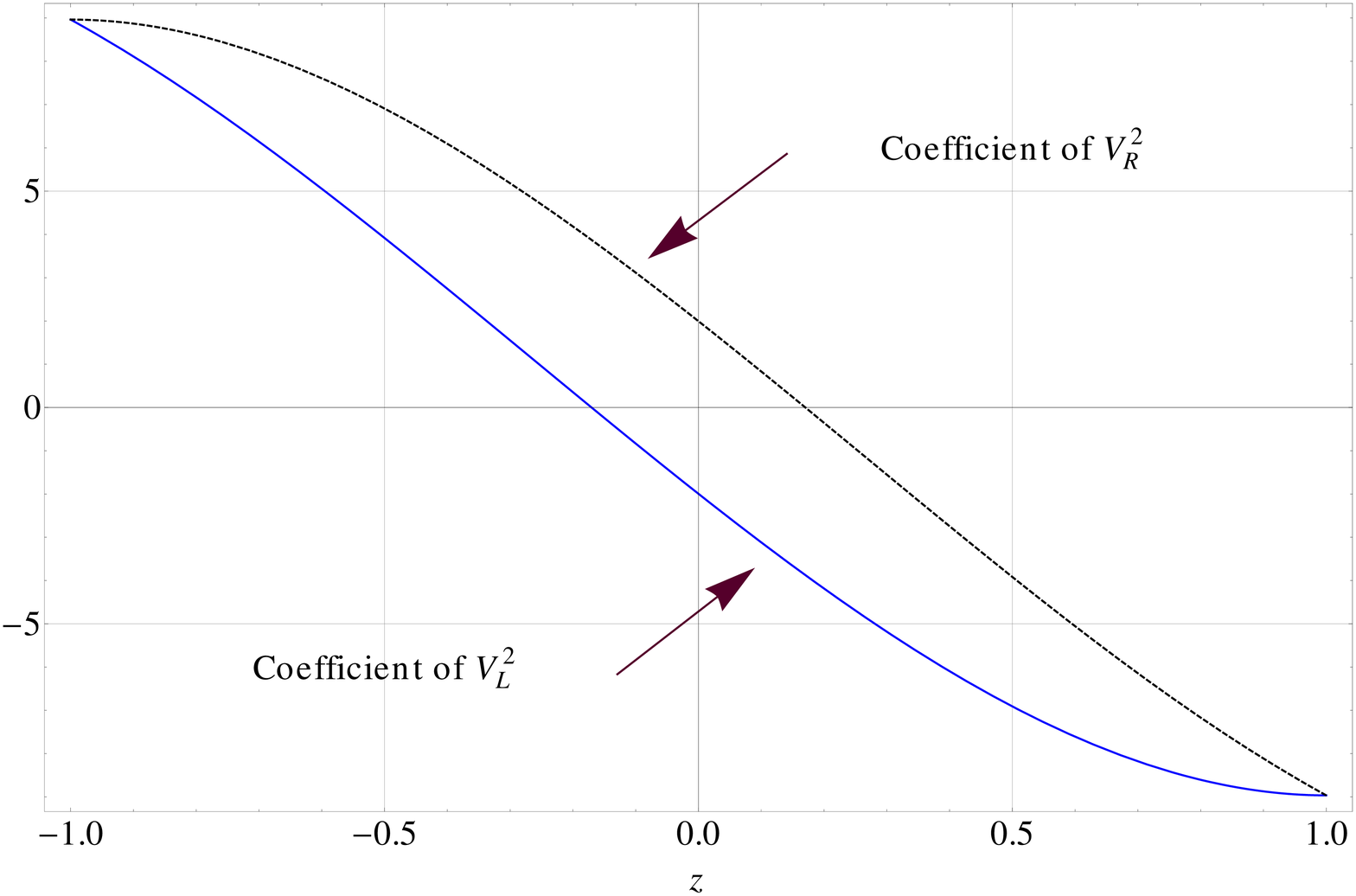}
\caption{Plot of the coefficients of $V_L^2$ and $V_R^2$ from eq. (\ref{eq:coeficientes1}) in terms of $z$.}
\label{figura2}
\end{minipage}
\hspace{0.02\textwidth}
\begin{minipage}[t]{0.50\textwidth}
\centering
\includegraphics[width=\textwidth]{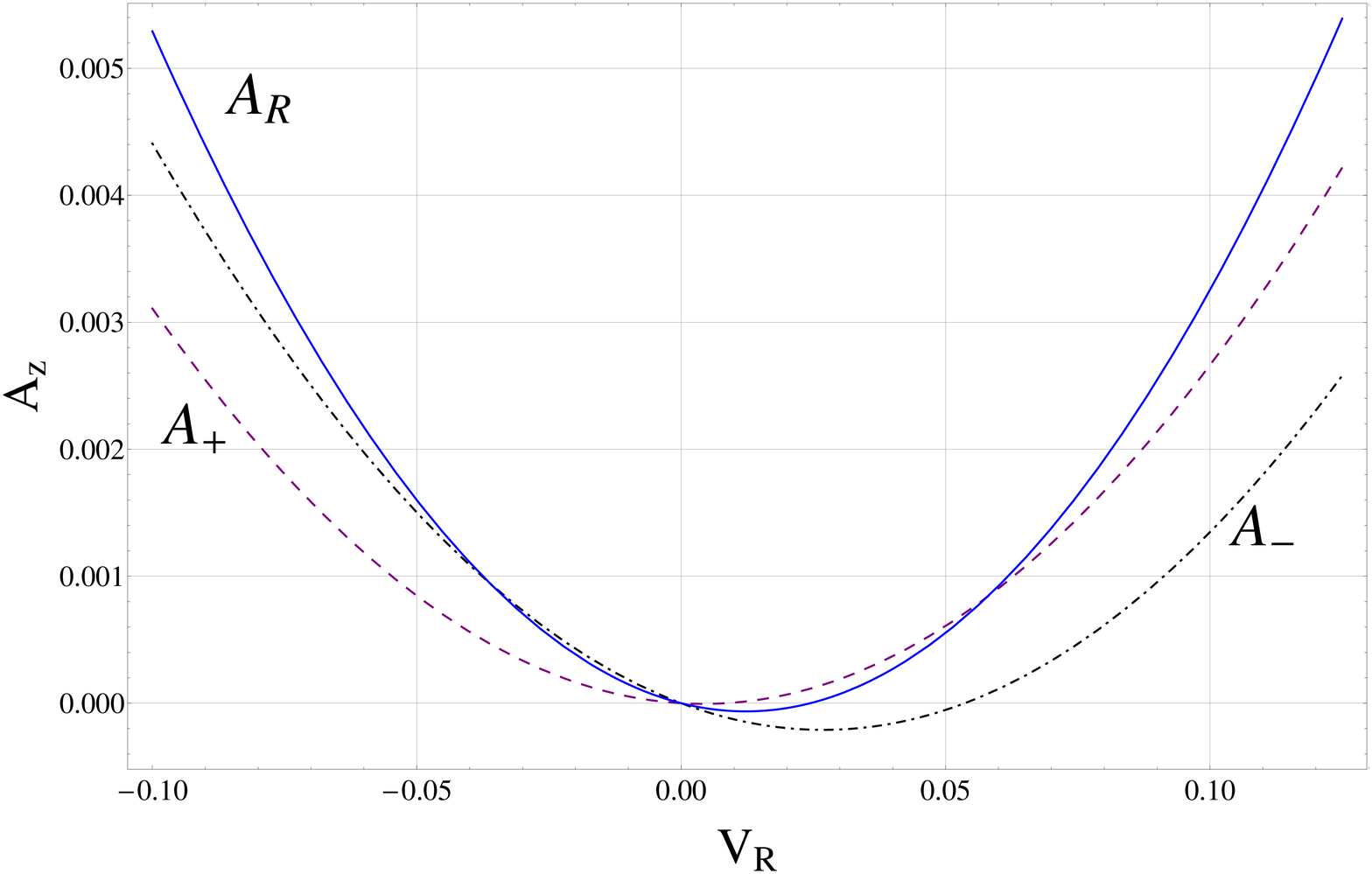}
\caption{Dependence on the $V_R$ coupling for  $A_R$ (blue-solid), eq. (\ref{eq:asimetriaR}), and for the usual $A_+$ (purple-dashed) and $A_-$ (black-dot-dashed) asymmetries, eq. (\ref{eq:asim1}).}
\label{fig:AsimetriaAz}
\end{minipage}
\end{figure}

The  helicity fractions, $F_0$ and $F_\pm$, can be computed in terms of the $tbW$ couplings given in eq. (\ref{gesdef}), and they can be found, for example, in ref. \cite{AguilarSaavedra:2006fy}. Performing the $\theta_l$ integration of eq. (\ref{eq:asim1}) for  a given value of  $z$, 
the numerator of the asymmetry $A_z$, in terms of  $V_{L}$ and $V_R$ is:
\begin{eqnarray}
&&V_L^2 \left(2.48 z^3 + 2.00 z^2 - 11.45 z- 2.00 \right) + 0.56 z V_L V_R \nonumber \\
&&+ V_R^2 \left(2.48 z^3 - 2.00 z^2 - 11.45 z + 2.00 \right).\label{eq:coeficientes1}
\end{eqnarray}
One can now choose the value of the $z$ parameter, within the range $(-1,1)$, to make zero the leading $V_L$ contribution. In this way we get the maximum sensitivity to $V_R$. The coefficient of $V_L^2$ and $V_R^2$ from eq. (\ref{eq:coeficientes1}), are plotted in figure \ref{figura2}. 
There, it can be seen that for $z = \displaystyle z_R = -0.17$  the coefficient of  $V_L^2$ cancels, leaving  the $V_R$ term, in eq. (\ref{eq:coeficientes1}), as the leading one. For this $z_R$
the new asymmetry $A_R\equiv A_Z(z=z_R)$  is proportional to $V_R$ and has the form:
\begin{equation}
A_R(V_L,V_R)=\frac{-0.01 V_L V_R + 0.48 V_R^2}{1.12 \left(V_L^2 + V_R^2\right) - 0.07 V_L V_R} \simeq - 0.01 V_R,
\label{eq:asimetriaR}
\end{equation}
where the last expression is the value of the  asymmetry taking  $V_L=1$ and assuming $|V_R|<<1$. As can be seen this asymmetry is directly proportional to $V_R$ so that a non-zero measurement  of it is a direct test of $V_R \neq 0$. 

Note that the leading contribution to the usual $A_\pm$ asymmetries comes from $V_L^2$  so that in order to be sensitive to $V_R$ one needs to subtract this SM central value.  In figure \ref{fig:AsimetriaAz} we show the  $V_R$  dependence of $A_R$ and also, for comparison, the dependence of $A_\pm$, with the $V_L^2$ leading contribution subtracted. As can be seen there $A_R$ may allow more precise bounds on $V_R$. Besides, it is directly proportional to $V_R$ and eliminates other uncertainties that the $V_L$ dependence in $A_\pm$ may introduce.

 For polarized top it is possible to define asymmetries with respect to the normal and transverse spin directions. These were studied in ref. \cite{AguilarSaavedra:2010nx}, but they are not sensitive to the $V_R$ coupling so we are not going to consider them in our analysis.

\subsection{Observables in the top rest frame}

 Angular distribution of the decay products for the weak process $t\rightarrow W^+ b$ carries information about the spin of the decaying top. Then,  angular asymmetries can be built to test the Lorentz structure of the vertex.  We will follow the same procedures described previously, in order to optimize the sensitivity of the observables  to $V_R$ but, in this case, for the asymmetries in the top rest frame.

For the top decay $t\rightarrow W^+\, b\rightarrow l^+\nu b,q\bar{q}'b$, the angular distribution of the product $X = l^+,\nu,q,\bar{q}',W^+,b$, in the top rest frame, is given by:
\begin{equation}
\frac{1}{\Gamma}\frac{d\Gamma}{d\cos\theta_X}=\frac{1}{2}\left(1+\alpha_X\, \cos\theta_X\right),
\end{equation}
where $\theta_X$ is the angle between the momentum of $X$ and the top spin direction, and $\alpha_X$ are the spin-analyzer powers, given in references \cite{AguilarSaavedra:2006fy,Grzadkowski:1999iq,Grzadkowski:2002gt,Godbole:2006tq,AguilarSaavedra:2010nx} in terms of the couplings shown in eq. (\ref{gesdef}). Then, an asymmetry can be defined as:
\begin{equation}
A_X^z\equiv  \frac{N(\cos\theta_X >z)-N(\cos\theta_X <z)}{N(\cos\theta_X >z)+N(\cos\theta_X <z)}=\frac{1}{2}\left[\alpha_X(1-z^2)-2z\right].\label{eq:asim2}
\end{equation}
For $z=0$, one gets the usual forward-backward asymmetry:
\begin{equation}
A_X\equiv  \frac{N(\cos\theta_X >0)-N(\cos\theta_X <0)}{N(\cos\theta_X >0)+N(\cos\theta_X <0)}=\frac{\alpha_X}{2}.\label{eq:asimX}
\end{equation}
Sensitivity to $V_R$ and $V_L$ for this asymmetry have already been given for $X=l^+,\; b,\; \nu$ in the $t$-channel single top production in refs. \cite{AguilarSaavedra:2008zz,AguilarSaavedra:2010nx}.  In order to define a new asymmetry directly proportional to the $V_R$ coupling one can again extract the SM leading contribution, given by the $V_L^2$ term in the $A_X^z$ asymmetry, and make it zero. For the $X=$ $l$, $\nu$, $b$ cases, the eq. (\ref{eq:asim2}) is:
\begin{eqnarray}
A_l^z&=&\frac{-1}{2\left(V_L^2+ V_R^2-0.06 V_L V_R\right)}\times\nonumber\\
&&\Bigg[V_L^2\left(-\frac{1}{2} z^2-z+\frac{1}{2}\right)+V_R^2 \left(-0.16 z^2-z+0.16\right)\nonumber\\
&&+0.06V_L V_R \left(\frac{1}{2}z^2+z-\frac{1}{2}\right)\Bigg]\label{eq:A_lz},
\end{eqnarray}
\begin{eqnarray}
A_b^z&=&\frac{0.20}{V_L^2+V_R^2-0.06 V_L V_R}\times\big[V_L^2\left( z^2-4.93 z-1\right) \nonumber\\
&&+V_R^2\left(-z^2-4.93 z+1\right)+0.31z V_L V_R \big],\label{eq:A_bz}
\end{eqnarray}
\begin{eqnarray}
A_\nu^z&=&\frac{0.16}{V_L^2+V_R^2-0.06 V_L V_R}\times\big[V_L^2\left(z^2-6.30 z-1\right)\nonumber\\
 &&+V_R^2\left(3.15 z^2-6.30 z-3.15\right)\nonumber\\
&&+V_L V_R\left(-0.19z^2+0.39 z+0.19\right)\big].\label{eq:A_nuz}
\end{eqnarray}
\begin{figure}[ht]
\centering
\begin{minipage}[t]{0.47\textwidth}
\centering
\includegraphics[width=\textwidth]{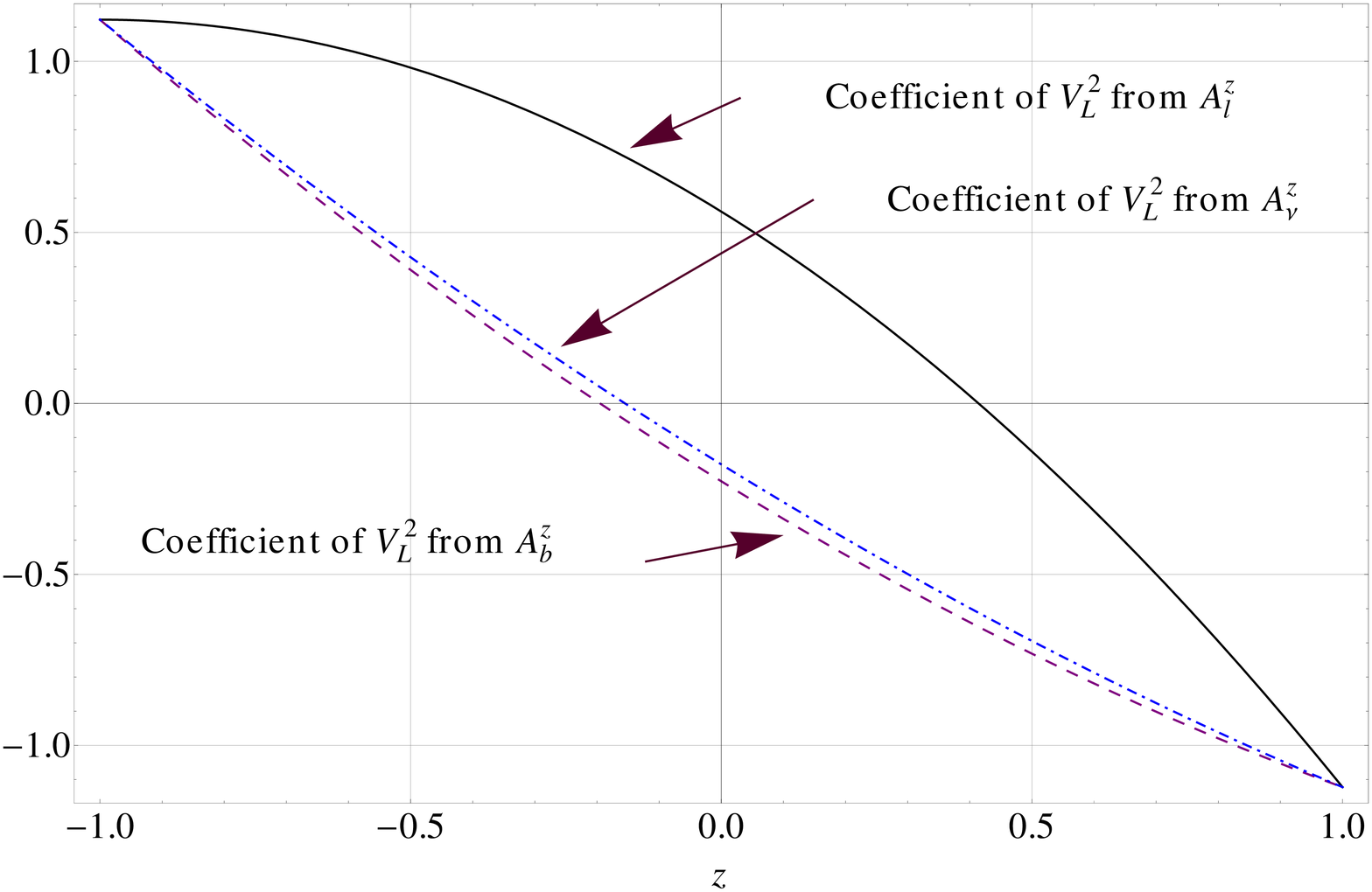}
\caption{Plot of the coefficients of $V_L^2$, in terms of $z$, from eqs. (\ref{eq:A_lz}), (\ref{eq:A_bz}) and (\ref{eq:A_nuz}).}
\label{fig:coeficientes2}
\end{minipage}
\hspace{0.02\textwidth}
\begin{minipage}[t]{0.47\textwidth}
\centering
\includegraphics[width=\textwidth]{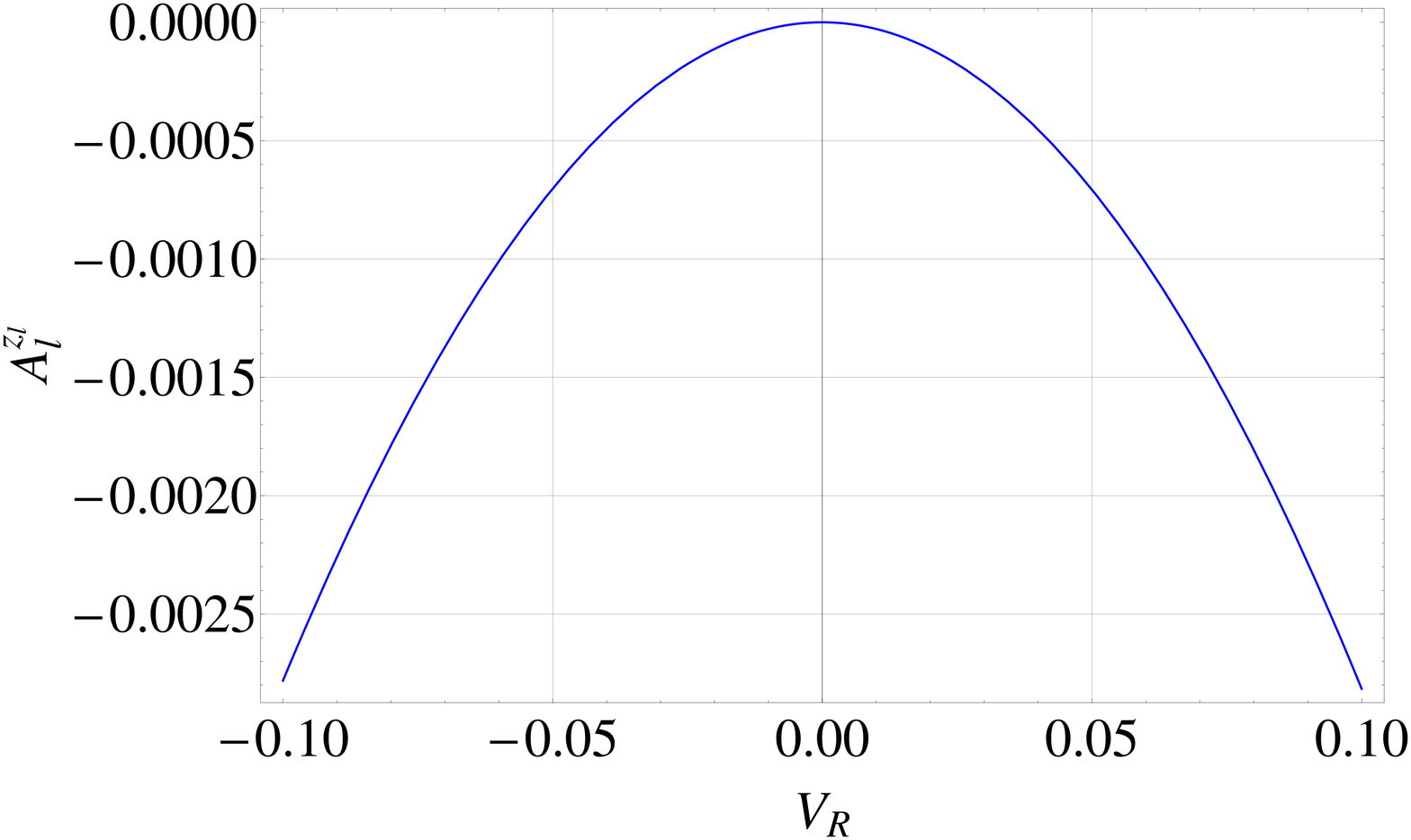}
\caption{Dependence on the $V_R$ coupling for the $A_l^{z_l}$ asymmetry, eq. (\ref{eq:a_l}).}
\label{fig:asimetriaAl}
\end{minipage}
\end{figure}

Figure \ref{fig:coeficientes2} shows the behavior of the $V_L^2$ coefficient from the $A_l$, $A_b$ and $A_\nu$ asymmetries. We can choose $z$ in order to make the leading coefficients of eqs. (\ref{eq:A_lz}), (\ref{eq:A_bz}) and (\ref{eq:A_nuz}) to be zero.  These  $z$ values are:
\begin{equation}
z_l=\sqrt{2}-1,\quad z_b=-0.20,\quad z_\nu=-0.16,
\end{equation}
and then, the new asymmetries are:
\begin{eqnarray}
A_l^{z_l}&=&\frac{-0.28 V_R^2}{V_L^2+ V_R^2-0.06 V_L V_R}  \simeq -0.28 V_R^2,  \label{eq:a_l}\\
A_b^{z_b}&=&\frac{0.39 V_R^2-0.01 V_L V_R}{V_L^2+ V_R^2-0.06 V_L V_R}  \simeq -0.01 V_R, \label{eq:a_b}  \\
A_\nu^{z_\nu}&=&\frac{0.02 V_L V_R-0.33 V_R^2}{V_L^2+ V_R^2-0.06 V_L V_R} \simeq 0.02 V_R, \label{eq:a_nu}
\end{eqnarray}
where in the last expression we show the value of the asymmetries for $V_L=1$ and assuming $|V_R|<<1$. From eq. (\ref{eq:A_lz}) one can see that unfortunately, for the $A_l$ asymmetry, the same value of $z$ that cancels the coefficient of $V_L^2$ also cancels the $V_L V_R$ term, in such a way that a poorer sensitivity to the coupling will be expected from this observable because the surviving term is $V_R^2$ instead of $V_L V_R$.
\begin{figure}[ht]
\centering
\begin{minipage}[t]{0.47\textwidth}
\centering
\includegraphics[width=\textwidth]{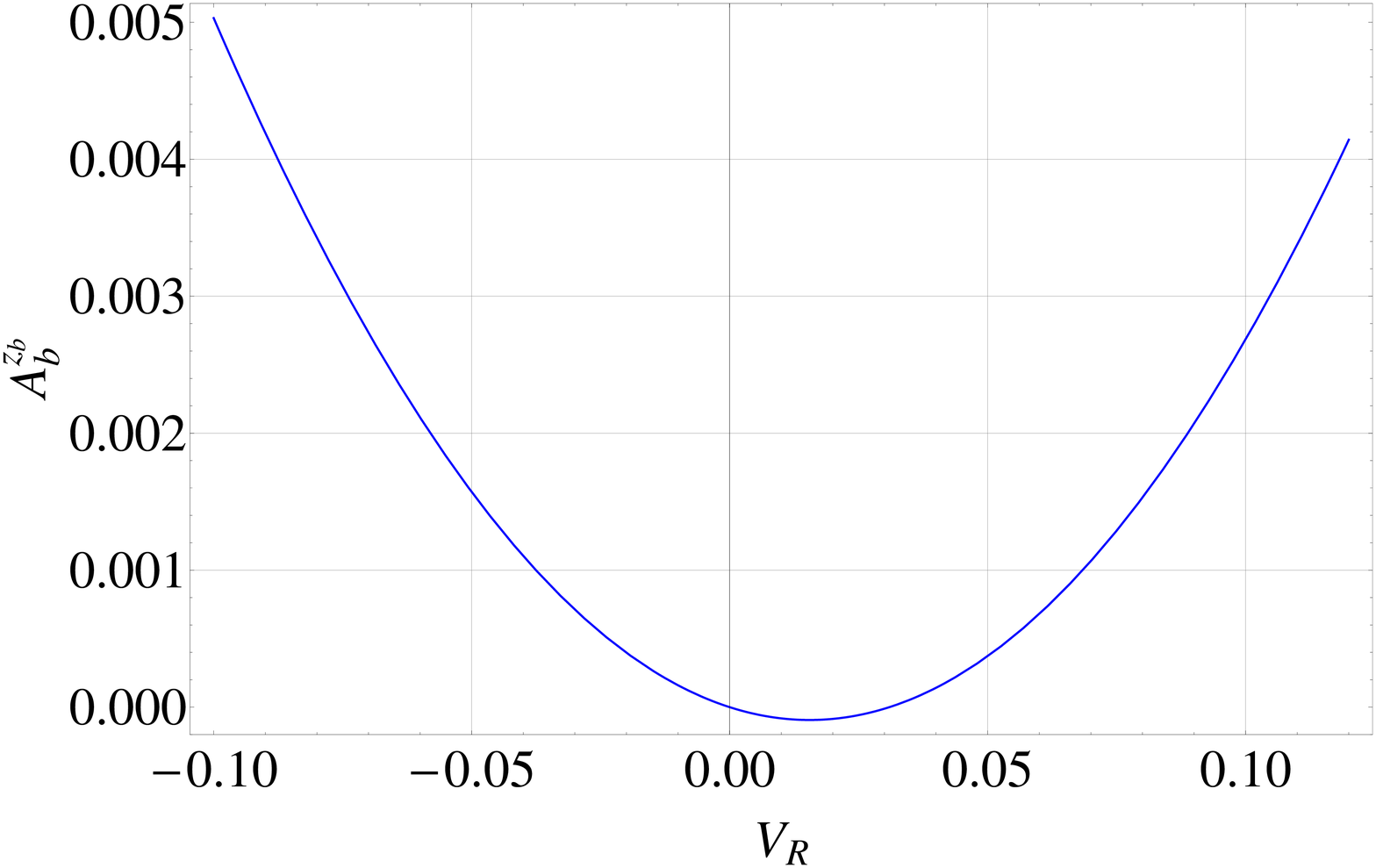}
\caption{Dependence  on the $V_R$ coupling  for the $A_b^{z_b}$ asymmetry, eq. (\ref{eq:a_b})}
\label{fig:asimetriaAb}
\end{minipage}
\hspace{0.02\textwidth}
\begin{minipage}[t]{0.47\textwidth}
\centering
\includegraphics[width=\textwidth]{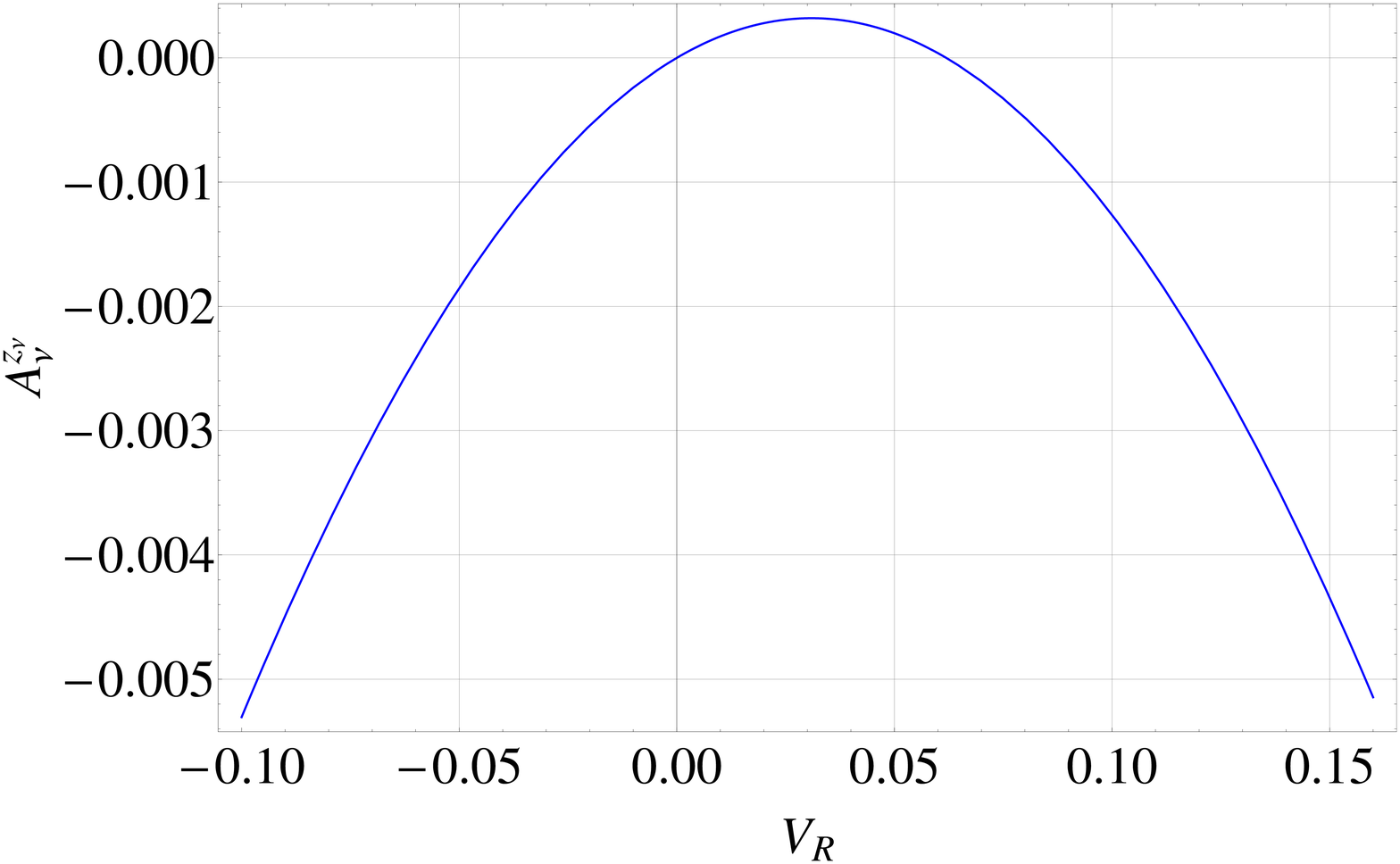}
\caption{Dependence  on the $V_R$ coupling for the  $A_\nu^{z_\nu}$ asymmetry, eq. (\ref{eq:a_nu}).}
\label{fig:asimetriaAnu}
\end{minipage}
\end{figure}

In figures \ref{fig:asimetriaAl}, \ref{fig:asimetriaAb} and \ref{fig:asimetriaAnu} we show the dependence on $V_R$ for the new $A_l^{z_l}$, $A_b^{z_b}$ and $A_\nu^{z_\nu}$ observables. We have explicitly checked that the usual forward-backward asymmetries for the same decay product have a very similar behavior as the ones shown in the figures, once the leading $V_L^2$ contribution is removed. However,  the new observables again have the advantage that are proportional to $V_R$  so that their measurement may provide a direct bound or measurement of the $V_R$ coupling with no assumption on the value of the $V_L^2$ leading term.

\subsection{Spin correlations in \texorpdfstring{$t\bar{t}$}{tt} production}

The top--antitop spin correlations depend on the Lorentz structure of the $tbW$ vertex.  This structure can be studied  through the measurement of the angular distributions of the decay products for the $t\rightarrow W^+b$  and $\bar{t}\rightarrow W^-\bar{b}$ processes, that carry information on the top and antitop spin correlation terms.  In particular, using the notation of previous section, the double angular distribution -- of the decay products $X$, from top, and $\bar{X}'$, from antitop -- can be written as \cite{Mahlon:1995zn, Stelzer:1995gc}:
\begin{equation}
\frac{1}{\sigma}\frac{d\sigma}{d\cos\theta_X\; d\cos\theta_{\bar{X}'}}=\frac{1}{4}\left(1+C\,\alpha_X\, \alpha_{\bar{X}'}\,\cos\theta_X\, \cos\theta_{\bar{X}'}\right),
\end{equation}
where $\theta_X$ ($\theta_{\bar{X}'}$) is the angle between the momentum of the decay product $X$ ($\bar{X}'$) and the momentum of the top (antitop) quark in the $t\bar{t}$ center-of-mass frame, $\alpha_X$ and $\alpha_{\bar{X}'}$ are the spin analyzer powers of particles $X$ and $\bar{X}'$, respectively. $C$ is a coefficient that weights the spin correlation between the quark and the antiquark. Then, one can define the asymmetry
\begin{eqnarray}
A_{X\bar{X}'}^{z_1 z_2}&\equiv&\frac{1}{\sigma}\Bigg[
\int_{z_1}^1d(\cos\theta_X)\int_{z_2}^1d(\cos\theta_{\bar{X}'})\,\frac{d\sigma}{d\cos\theta_X\; d\cos\theta_{\bar{X}'}}\nonumber\\
&+&\int_{-1}^{z_1}d(\cos\theta_X)\int_{-1}^{z_2}d(\cos\theta_{\bar{X}'})\, \frac{d\sigma}{d\cos\theta_X\; d\cos\theta_{\bar{X}'}}\nonumber\\
&-&\int_{-1}^{z_1}d(\cos\theta_X)\int_{z_2}^1d(\cos\theta_{\bar{X}'})\,\frac{d\sigma}{d\cos\theta_X\; d\cos\theta_{\bar{X}'}}\nonumber\\
&-&\int_{z_1}^1d(\cos\theta_X)\int_{-1}^{z_2}d(\cos\theta_{\bar{X}'})\, \frac{d\sigma}{d\cos\theta_X\; d\cos\theta_{\bar{X}'}}\Bigg].\label{eq:azz}
\end{eqnarray}
\begin{figure}[ht]
\centering
\begin{minipage}[t]{0.47\textwidth}
\centering
\includegraphics[width=\textwidth]{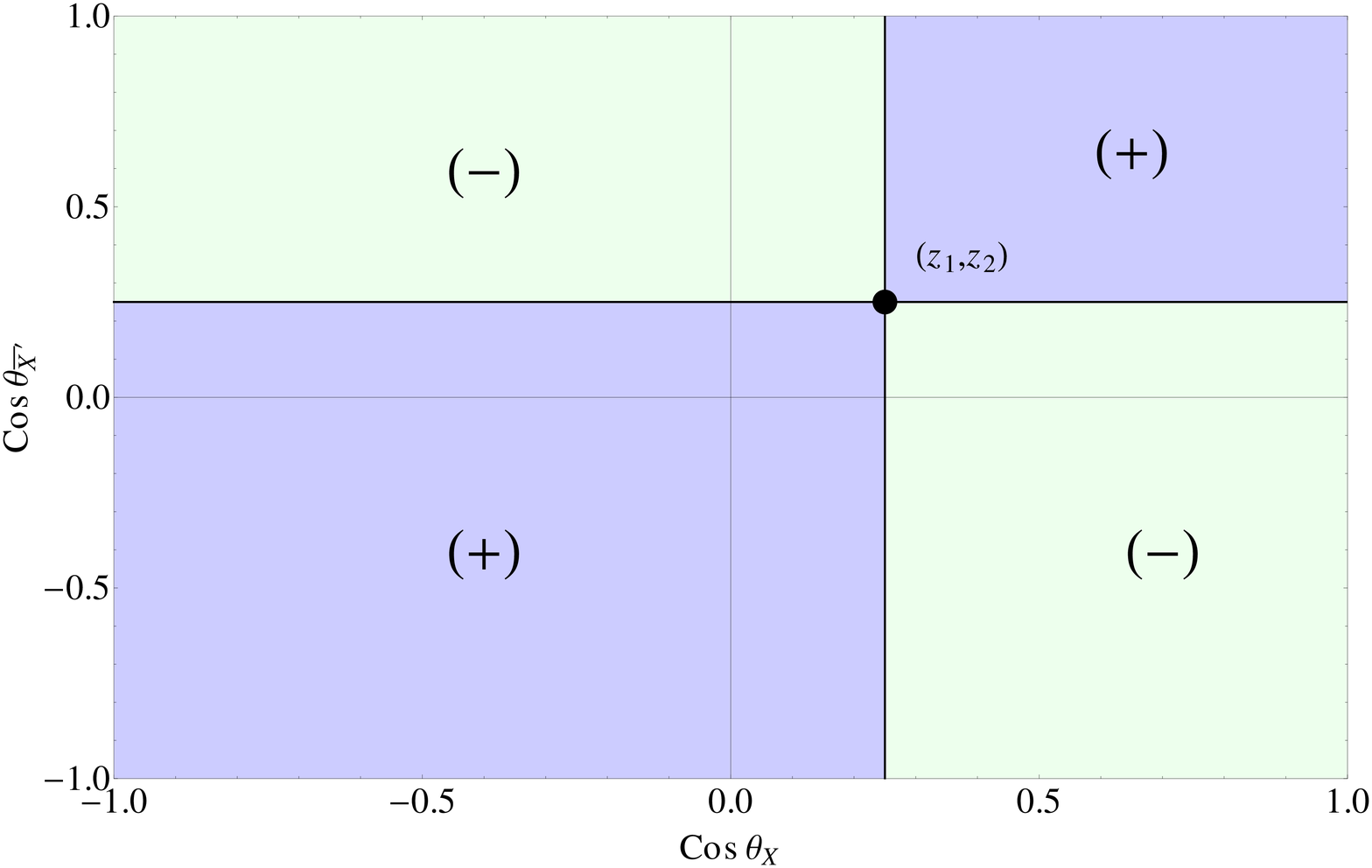}
\caption{Regions of integration and sign of the contribution to the spin correlation asymmetry $A_{X\,\bar{X}'}^{z_1\, z_2}$  defined in eq. (\ref{eq:azz}), depending on the point $(z_1,z_2$)}
\label{fig:areas}
\end{minipage}
\hspace{0.02\textwidth}
\begin{minipage}[t]{0.47\textwidth}
\centering
\includegraphics[width=\textwidth]{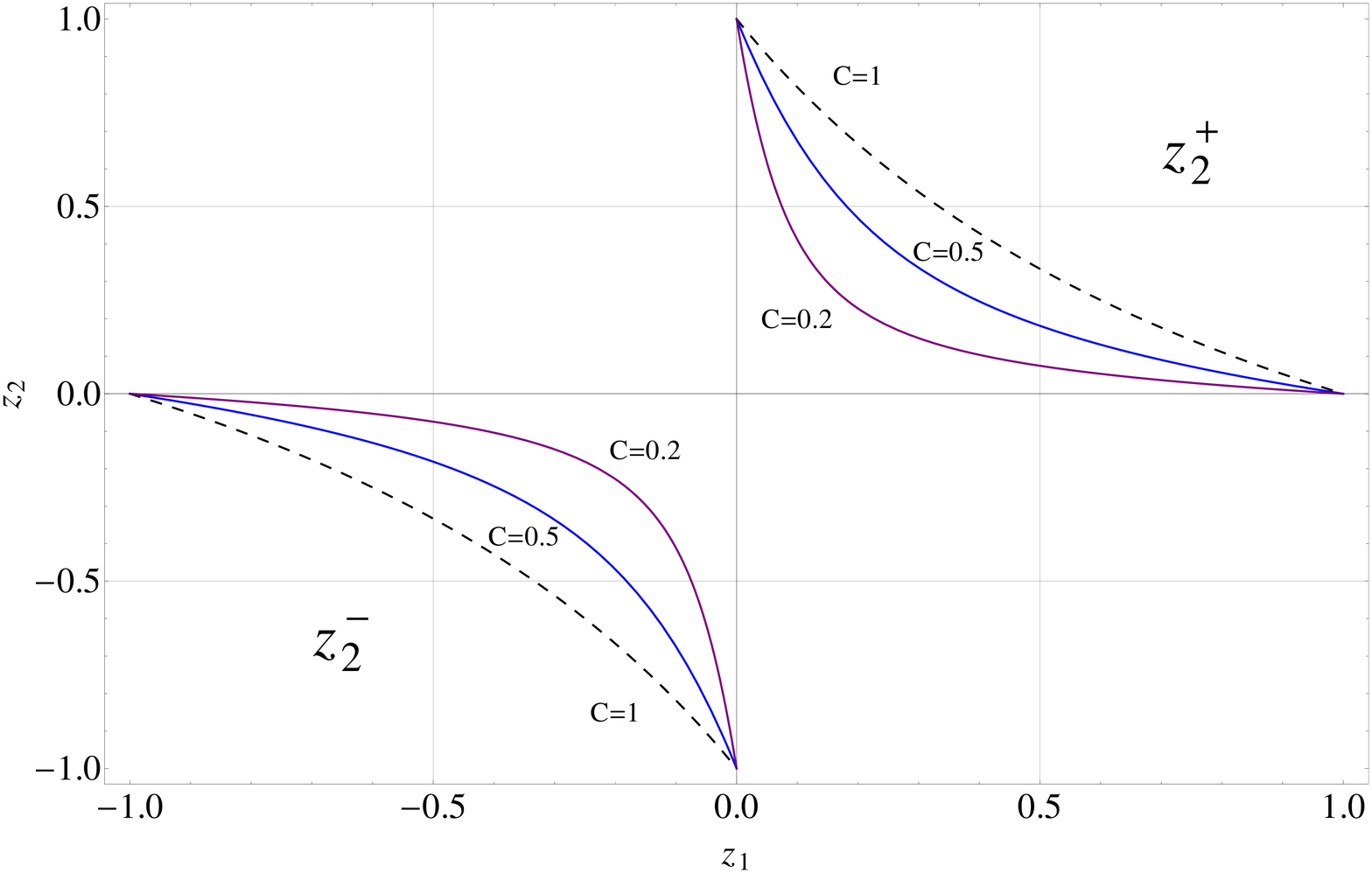}
\caption{Values of $z_1$ and $z_2$ that makes zero the $V_L$ leading terms of the $A_{ll'}^{z_1z_2}$ asymmetry, eq. (\ref{eq:z1z2}), for different values of the spin correlation coefficient $C$.}
\label{fig:z1z2}
\end{minipage}
\end{figure}
The different regions of integration for this asymmetry are plotted in figure \ref{fig:areas}. There, the number of  events on each of the regions is collected and the sign of their contribution to the asymmetry is also indicated. The particular case $z_1=z_2=0$ corresponds to the usual spin correlation asymmetry:
\begin{equation}
A_{X\, \bar{X}'}=\frac{N\left(\cos\theta_X\,\cos\theta_{\bar{X}'}>0\right)- N\left(\cos\theta_X\,\cos\theta_{\bar{X}'}<0\right)}
{N\left(\cos\theta_X\,\cos\theta_{\bar{X}'}>0\right)+ N\left(\cos\theta_X\,\cos\theta_{\bar{X}'}<0\right)}.\label{eq:Azz}
\end{equation}
 For the $A_{X\bar{X}'}^{z_1 z_2}$ asymmetry, eq. (\ref{eq:azz}), there is a range of values of $z_1$ and $z_2$ that cancel the leading $V_L^4$ contribution,  in such a way that  the observable becomes proportional to the $V_R$ coupling. For the $X=l$, $\bar{X}'=l'$ case, these values satisfy the relation: 
\begin{equation}
z_2^\pm=\frac{-2z_1\pm\sqrt{C^2 (1-z_1^2)^2+4z_1^2}}{C (1-z_1^2)}.\label{eq:z1z2}
\end{equation}
Similarly to what happened for $A_l^z$, in eq. (\ref{eq:A_lz}), for values of $z_1$ and $z_2$ satisfying the previous equation, the coefficient of $V_L^3 V_R$  in the asymmetry also accidentally cancels. Then, it remains the  $V_R^2$ coupling as the leading contribution. Solutions to eq. (\ref{eq:z1z2}) are plotted in figure \ref{fig:z1z2} where it can be seen that the values of  $z_1$ and $z_2$ are restricted to be in the first quadrant (for the $z_2^+$ solution) or in the third one (for the $z_1^-$ solution). In addition to the trivial solutions of eq. (\ref{eq:z1z2}), i.e. $z_1=\pm 1,z_2=0$, that reproduces the usual $A_{ll'}$ asymmetry, one can improve the computation  by finding the values of $z_1$ and $z_2$ that, satisfying eq. (\ref{eq:z1z2}), also maximize the coefficient of the leading $V_R^2$ term in eq.(\ref{eq:azz}), namely:
\begin{equation}
2.00\, V_R^2 \left[z_1 z_2 +0.08\, C \left(z_1^2 \left(1- z_2^2\right)+ z_2^2-1\right)\right].
\end{equation}
This can be trivially done and the result is 
\begin{equation}
 z_1=z_2=z_{ll'}\equiv \pm\sqrt{1+\frac{2}{C}-\frac{2}{C}\sqrt{1+C}},\label{eq:zllmax}
\end{equation}
that correspond to the intersection of the curves of figure \ref{fig:z1z2} with the diagonal line of the first and third quadrants.
\begin{figure}[ht]
\centering
\begin{minipage}[t]{0.47\textwidth}
\centering
\includegraphics[width=\textwidth]{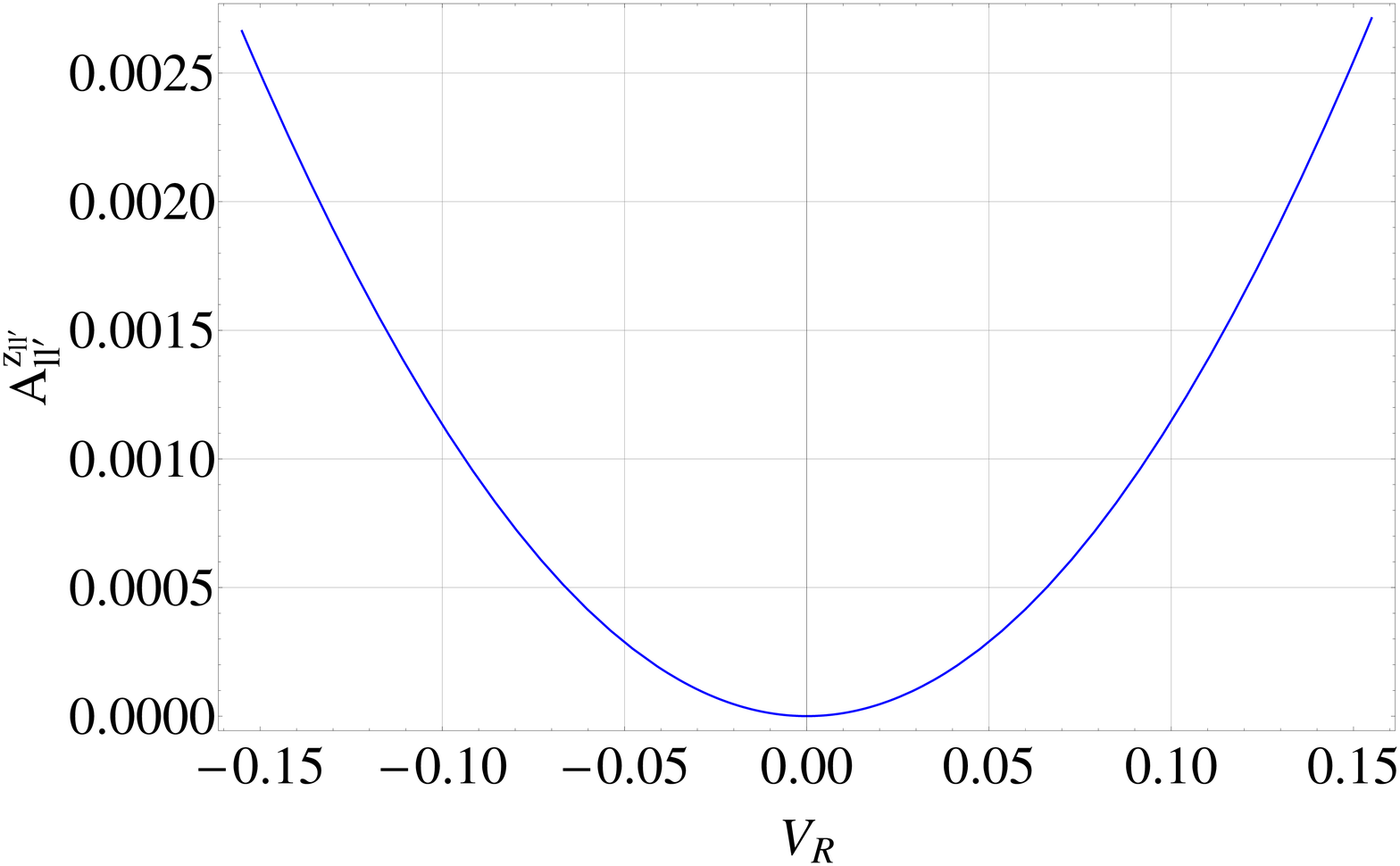}
\caption{Dependence on $V_R$ coupling for $A_{ll'}^{z_1z_2}$ in eq. (\ref{eq:azz}), for $C=0.4$ and $z_1=z_2=z_{ll'}=0.29$.} 
\label{fig:deltaAll}
\end{minipage}
\hspace{0.02\textwidth}
\begin{minipage}[t]{0.47\textwidth}
\centering
\includegraphics[width=\textwidth]{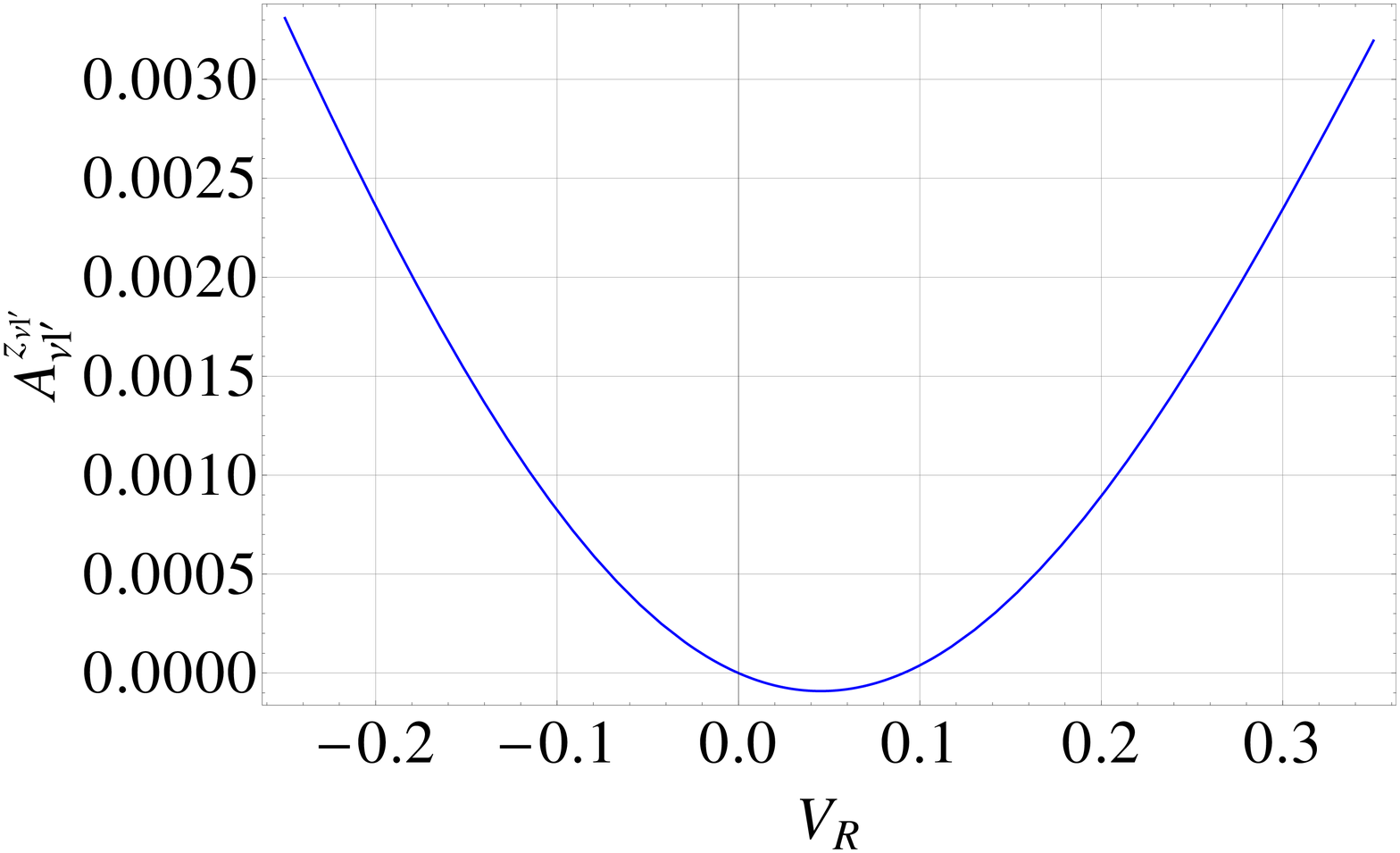}
\caption{Dependence on the $V_R$ coupling for $A_{\nu l'}^{z_1z_2}$, eq. (\ref{eq:azz}), for $C=0.4$ and $z_1=-z_2=z_{\nu l'}=0.17$.} 
\label{fig:AsimetriaAnul}
\end{minipage}
\end{figure}
These values of $z_1$, $z_2$  give a maximum sensitivity of the $A_{ll'}^{z_1z_2}$ asymmetry to the $V_R$ coupling. 

In figure \ref{fig:deltaAll} we show the dependence on the $V_R$ coupling  for the $A_{ll'}^{z_1z_2}$ asymmetry in eq. (\ref{eq:azz}), for $C=0.4$ \cite{AguilarSaavedra:2010nx} ,  and  $z_1=z_2=z_{ll'}=0.29$, given by eq. (\ref{eq:zllmax}). 

 The same procedure followed here can be used for other decay products  of the $W^\pm$. For $t\bar{t} \rightarrow l\,\nu\, b\, \bar{\nu}\, l'\,\bar{b}$ final state, the values of $z_1$ and $z_2$ that cancels the coefficient of the leading $V_L^4$ term in the  $A_{\nu l'}^{z_1 z_2}$ asymmetry ($X=\nu$, $\bar{X}'=l'$ in eq. (\ref{eq:azz})),  satisfy a quadratic equation
\begin{equation}
z_1z_2=-\beta(1-z_1^2)(1-z_2^2),\quad \beta=0.08C,
\end{equation}
and the solution is 
\begin{equation}
  z_2^\pm=\frac{1}{2\beta(1-z_1^2)}\left[z_1\pm\sqrt{z_1+4\beta^2(1-z_1)^2}\right].
\label{eq:z1z2nul}
\end{equation}
Here, $-1\leq z_1\leq 0$ for the $z_2^+$ solution and $0\leq z_1\leq 1$  for the $z_2^-$ solution  of eq. (\ref{eq:z1z2nul}),  in order to satisfy $|z_2^\pm|<1$. In that case, these set of $z_{1,2}$ values do not cancel the $V_R$ term of the asymmetry, that remains as the leading one:
\begin{equation}
-0.12\left[z_1 z_2-\frac{C}{4}(1-z_1^2)(1-z_2^2)\right] V_R.
\end{equation}
One can easily find  the values that maximize this coefficient and simultaneously verify eq. (\ref{eq:z1z2nul}):
\begin{equation}
z_1=-z_2=z_{\nu l'}\equiv\pm\sqrt{1+\frac{1}{2\beta}-\frac{1}{2\beta}\sqrt{1+2\beta}} 
\label{eq:znul}
\end{equation}
In figure \ref{fig:AsimetriaAnul} we show the dependence of the spin correlation asymmetry $A_{\nu l'}^{z_{\nu l'}z_{\nu l'}}$ on $V_R$, for $z_1$, $z_2$ given by eq. (\ref{eq:znul})  with $C=0.4$.  Note that the sensitivity of both asymmetries shown in figures  \ref{fig:deltaAll} and  \ref{fig:AsimetriaAnul} is rather similar. 

\begin{figure}[ht]
\centering
\begin{minipage}[t]{0.47\textwidth}
\centering
\includegraphics[width=\textwidth]{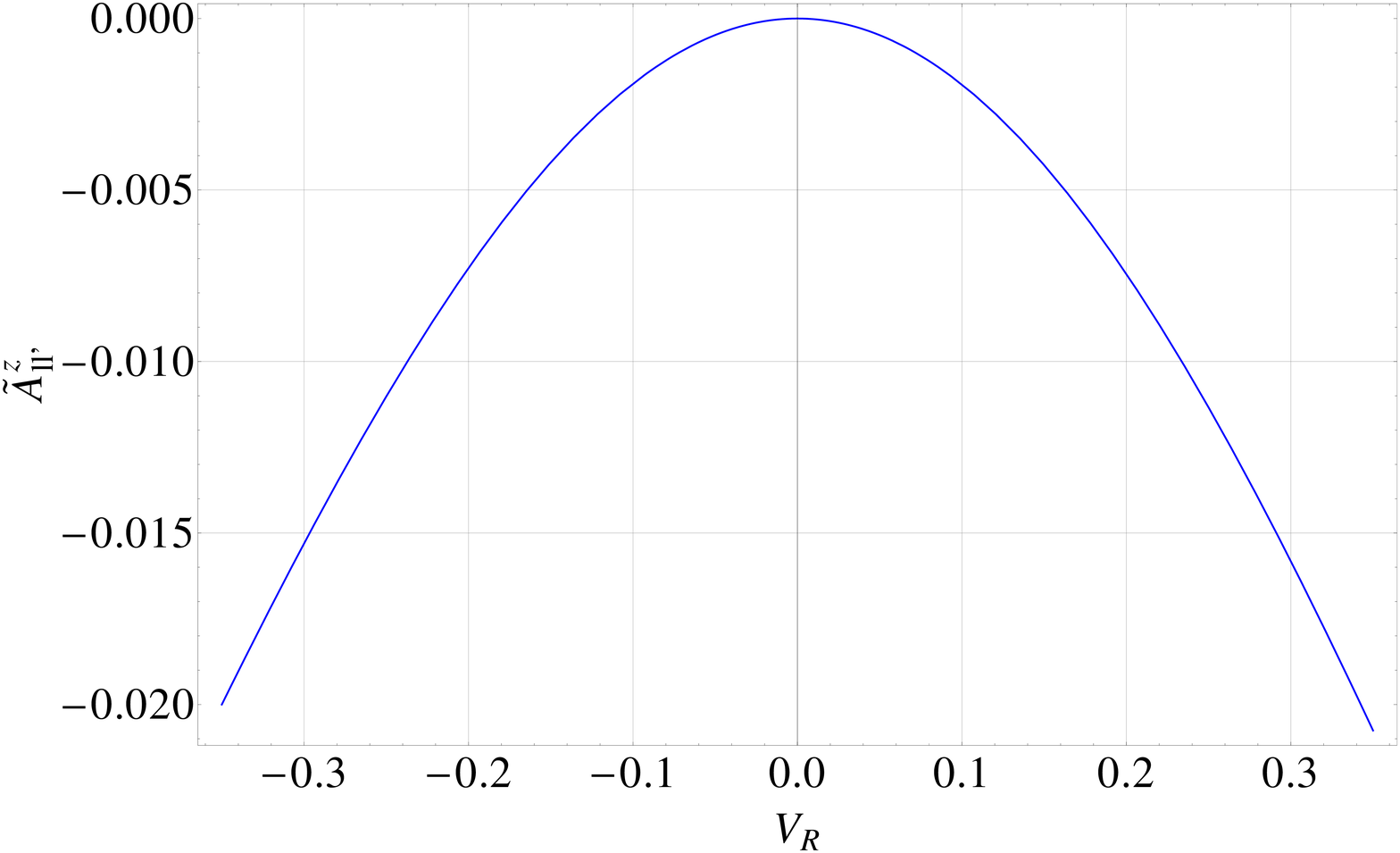}
\caption{Dependence on the $V_R$ coupling for the $\tilde{A}_{ll'}^{z}$ asymmetry, eq. (\ref{eq:tildeaz}), for $D=-0.29$ and $z=\tilde{z}_{ll'}$ given by eq. (\ref{eq:tildez}).}
\label{fig:Asimetria_tilde_ll}
\end{minipage}
\hspace{0.02\textwidth}
\begin{minipage}[t]{0.47\textwidth}
\centering
\includegraphics[width=\textwidth]{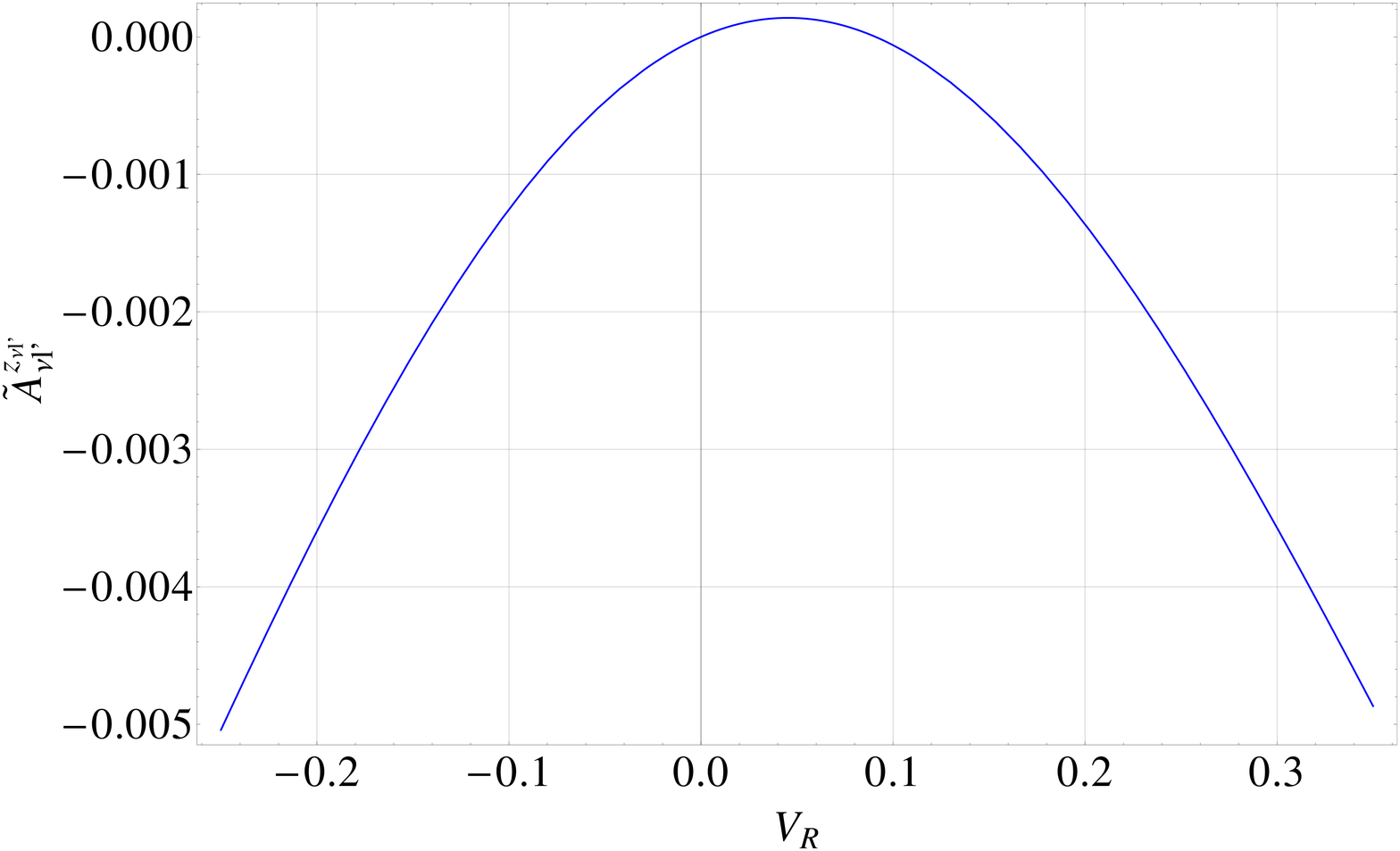}
\caption{Dependence on the $V_R$ coupling for the $\tilde{A}_{\nu l'}^{z}$ asymmetry, eq. (\ref{eq:tildeanul}), for $D=-0.29$ and $z=z=\tilde{z}_{\nu l'}$ given by eq. (\ref{eq:tildeznul}).} 
\label{fig:Asimetria_tilde_nul}
\end{minipage}
\end{figure}

Another spin correlation considered in the literature is the angular distribution of the top (antitop) decay products  defined as \cite{Bernreuther:2004jv}:
\begin{equation}
\frac{1}{\sigma}\frac{d\sigma}{d\cos\varphi_{X\bar{X}'}}=\frac{1}{4}\left(1+D\,\alpha_X\, \alpha_{\bar{X}'}\, \cos\varphi_{X\bar{X}'}\right),
\end{equation}
where $\varphi_{X\bar{X}'}$ is the angle between the momentum of the $X$ particle in the $t$ rest frame, and that of the $X'$ one, in the $\bar{t}$ rest frame. D is the  spin correlation coefficient. Then, one can construct the following asymmetry:
\begin{equation}
\tilde{A}_{X\bar{X}'}^z\equiv\frac{N(\cos\varphi_{X\bar{X}'}>z)-N(\cos\varphi_{X\bar{X}'}<z)}{N(\cos\varphi_{X\bar{X}'}>z)+N(\cos\varphi_{X\bar{X}'}<z)}
=\frac{1}{2}\left(\alpha_X\, \alpha_{\bar{X}'}(1-z^2)-2z\right).\label{eq:barAzz}
\end{equation}
For $z=0$ one gets the usual forward-backward spin correlation asymmetry 

\begin{equation}
\tilde{A}_{X\bar{X}'}=\frac{1}{2}D\, \alpha_X\, \alpha_{\bar{X}'}.\label{eq:tildeAz}
\end{equation}
For both $W^\pm$ leptonic decays, $X=l$ and $\bar{X}'=l'$,  and following similar procedures as in the previous sections, we can find the value of $z$ that cancels the $V_L^4$ leading term  in eq. (\ref{eq:barAzz}):
\begin{equation}
z=\tilde{z}_{ll'}\equiv\frac{1}{D}\left(1-\sqrt{1+D^2}\right).\label{eq:tildez}
\end{equation}
This value also cancels the $V_R$ term so that the $V_R^2$ contribution is the dominant one.  Then, for $D=-0.29$ \cite{AguilarSaavedra:2010nx}, the $\tilde{A}_{ll'}^z$ asymmetry is:
\begin{equation}
\tilde{A}_{ll'}^{\tilde{z}_{ll'}}=\frac{-0.19 V_L^2 V_R^2+0.01 V_L V_R^3-0.13 V_R^4}{\left(V_L^2-0.06 V_L V_R+V_R^2\right)^2} \simeq -0.19 V_R^2 \label{eq:tildeaz}
\end{equation}
where again, in the last term, we show the asymmetry for $V_L=1$ and $|V_R|<<1$. 
This asymmetry can be seen in figure \ref{fig:Asimetria_tilde_ll}.

Analogously, for $W$ leptonic decays with neutrino-lepton final state, $X=\nu$, $\bar{X}'=l'$,  the $z$-value that cancels the $V_L^4$ term in eq. (\ref{eq:barAzz}) is 
\begin{equation}
z=\tilde{z}_{\nu l'}=-\frac{3.15}{D}\left(1-\sqrt{1+\left(\frac{D}{3.15}\right)^2}\right),
\label{eq:tildeznul}
\end{equation}
and then,  the $\tilde{A}_{\nu l'}^z$ asymmetry, for $D=-0.29$, is:

\begin{equation}
\tilde{A}_{\nu l'}^{\tilde{z}_{\nu l'}}=\frac{0.006 V_L^3 V_R-0.068 V_L^2 V_R^2+0.006 V_L V_R^3}{(V_L^2 - 0.062 V_L V_R + V_R^2)^2}\simeq 0.006 V_R. \label{eq:tildeanul} 
\end{equation}
The last term in this equation is given for $V_L=1$ and $|V_R|<<1$. This asymmetry is depicted in figure \ref{fig:Asimetria_tilde_nul}. Note that in this case the sensitivity to $V_R$ is lower than the one obtained in some of the previous observables due to the small coefficients in the numerator of eq. (\ref{eq:tildeanul}).

 We  checked that the  expressions of the observables defined for the $V_R$ coupling in this section show a behavior comparable to the one obtained from the expressions of the observables defined in the literature,  once the $V_L$ leading contribution is removed. Moreover, the observables defined here have the advantage of being directly proportional to the $V_R$ term that we want to test.

\section{Conclusions}
We  computed the SM one-loop QCD and electroweak contribution to $V_R$. Due to accidental cancellations between the diagrams, the leading contribution is mainly coming from QCD, it is real and of the order of $10^{-3}$. Any measurement of observables that may lead to $V_R$ above to $10^{-3}$ should be interpreted as new physics effects. 

We also have proposed  new observables that may provide a direct measurement of the right coupling. 
We found that for several angular asymmetries considered in the literature it is possible to define new observables, with an optimal choice of parameters, in such a way that they become a direct probe of $V_R$. These observables include angular asymmetries in the $W$ rest frame, angular asymmetries in the top rest frame and also spin correlations.  All the new observables defined in this paper are proportional to $V_R$ and then suitable for a direct determination of this coupling. In some cases the behavior shown by the expressions for our observables,  presents a better sensitivity to $V_R$ (like it is shown in Figure 3). While the asymmetries usually considered in the literature have leading contributions from  $V_L$, the new asymmetries that we have studied  here have as a leading term the $V_R$ right coupling we are interested in. These asymmetries can be measured with LHC data, where a huge number of top events are being collected, in order to obtain a direct measurement on the standard model contribution to the right top quark.
 
\section{Acknowledgments}

This work has been supported, in part, by the Ministerio de Ciencia e Innovaci\'on, Spain, under grants FPA2011-23897 and FPA2011-23596; by Ministerio de Economía y Competitividad, Spain, under grants FPA2014-54459-P and SEV-2014-0398; by Generalitat Valenciana, Spain, under grant PROMETEOII2014-087; and by CSIC and Pedeciba, Uruguay.
 
 \begin{appendices}
\section{Diagram contributions}
\label{app:AppendixA}
Using the following definitions for the denominators: 
\begin{eqnarray}
&&\hspace*{-1.5cm} A_Z= x^2 \left[\left((y-1)  r_b^2+1\right) y-  r_w^2 (y-1)\right]-r_z^2 (x-1), \\ 
&&\hspace*{-1.5cm} B_Z= x \left\{\left[\left(x (y-1)+1\right)  r_b^2+x-1\right]y -r_z^2 (y-1)\right\}- r_w^2 (x-1) \left[x (y-1)+1\right], \\ 
&&\hspace*{-1.5cm} C_Z= (x-1) (x y-1)  r_b^2-r_w^2 (x-1) x (y-1)\ +r_z^2 xy+x (y-1) (x y-1),\\ 
&& \hspace*{-1.5cm}\left\{A_\gamma,B_\gamma,C_\gamma\right\}
=\left\{A_Z,B_Z,C_Z\right\}(r_z\rightarrow 0),\\ 
&& \hspace*{-1.5cm}\left\{A_H,B_H,C_H\right\}
=\left\{A_Z,B_Z,C_Z\right\}(r_z\rightarrow r_h).\label{anal:tgw}
\end{eqnarray}
with
\begin{equation}
r_x\equiv \frac{m_x}{m_t}
\label{rx}
\end{equation}
and taking the usual definition for the SM couplings
\begin{equation}
a_t=-a_b=1, \quad v_b=-1+\frac{4s_w^2}{3} \quad \mbox{and} \quad v_t=1-\frac{8s_w^2}{3},
\end{equation}
the expression for the contribution of each diagram is given by:
\begin{equation}
I^{tZW}=
-\frac{1}{32  \pi  s_w^2}\times
\int_0^1dx \int_0^1dy\;
\frac{2x^2y\left[v_t(1+2xy)-a_t(5-2xy)\right]}
{A_Z},\label{tZW}
\end{equation}
\begin{equation}
I^{t \gamma W}=
-\frac{1}{8  \pi }\, Q_t \times
\int_0^1dx \int_0^1dy\;
\frac{2  x^2y(1+2xy)}
{A_\gamma},\hfill
\end{equation}
\begin{equation}
I^{t H W} = 0,
\end{equation}
\begin{eqnarray}
I^{t w_0 w}+ I^{t H w}&=&
\frac{1}{16  \pi  s_w^2}\frac{1}{r_w^2} \times\int_0^1dx \int_0^1dy\;\Bigg\{
-\frac{x^3 y\left[1+y-r_b^2(1-y)\right]}
{A_Z}\nonumber\\
&&\hspace*{2.cm}+(1-rb^2)\;
\frac{x^3y (1-y)}
{A_H}+x\log\frac{A_H}{A_Z}\Bigg\},
\label{div1}\end{eqnarray}
\begin{equation}
I^{t Z w}=
\frac{1}{32\pi c_w^2} \times
\int_0^1dx \int_0^1dy\;
\frac{2x^2y(a_t-v_t)}
{A_Z},
\end{equation}
\begin{equation}
I^{t \gamma w}=
-\frac{1 }{8 \pi }\,Q_t \times
\int_0^1dx \int_0^1dy\;
\frac{2 x^2 y}
{A_\gamma},
\end{equation}
\begin{equation}
I^{b W Z}=
\frac{1}{32  \pi   s_w^2} \times
\int_0^1dx \int_0^1dy\; \frac{2 x^2y\left[v_b(1+2xy)-a_b(5-2xy)\right]}{B_Z},
\end{equation}
\begin{equation}
I^{b W \gamma}=
\frac{1}{8  \pi   }\,Q_b \times
\int_0^1dx \int_0^1dy\;
\frac{2 x^2y(1+2xy)}
{B_\gamma},
\end{equation}
\begin{equation}
I^{b W H}= 0,
\end{equation}
\begin{eqnarray}
I^{b w w_0}+I^{b w H}&=&
\frac{1}{16  \pi s_w^2}\frac{1}{r_w^2}\times\int_0^1dx \int_0^1dy\; \Bigg\{\frac{
x^2 y\left[1-x-r_b^2\left(1-x(1-2y)\right)\right]}
{B_Z}\nonumber\\
&&\hspace*{2cm}-(1-r_b^2)
\frac{x^2 y(1-x)}
{B_H}+ x\log\frac{B_H}{B_Z}\Bigg\},
\label{div2}\end{eqnarray}
\begin{equation}
I^{b w Z}=
\frac{1}{32\pi  c_w^2}\times
\int_0^1dx \int_0^1dy\;
\frac{2x^2y(v_b-a_b)}
{B_Z},
\end{equation}
\begin{equation}
I^{b w\gamma}=
-\frac{1}{8  \pi }\,Q_b\times
\int_0^1dx \int_0^1dy\;
\frac{2 x^2 y}
{B_\gamma},
\end{equation}
\begin{eqnarray}
I^{Z t b}&=&
\frac{1}{32  \pi  c_w^2  s_w^2}\nonumber\\  &&\ns\ns\ns \times
\int_0^1dx \int_0^1dy\;
\frac{x\left[a_b(1+xy)-v_b(1-xy)\right]\left[a_t(1+xy)-v_t(1-xy)\right]}{C_Z},
\end{eqnarray}
\begin{equation}
I^{\gamma t b}=
\frac{1}{2  \pi  }\, Q_b\,  Q_t \times
\int_0^1dx \int_0^1dy\;
\frac{x(1-xy)^2}
{C_\gamma},\label{gtb}
\end{equation}
\begin{eqnarray}
I^{w_0 t b}+I^{Htb}&=&
\frac{1}{16  \pi s_w^2}\times\nonumber\\
&&\hspace*{-2cm}\int_0^1dx \int_0^1dy\; x^2(1-x)(1-y)\Bigg\{
\frac{-1}{C_Z}+\frac{1}{C_H}+\frac{x}{r_w^2}\log\frac{C_Z}{C_H}\Bigg\}.\label{Htb}
\end{eqnarray}

\section{Contribution of diagrams with a photon}
\label{app:AppendixB}

\begin{flalign}
&I_0^{t\gamma W} = \frac{-2}{r_b}\Bigg\{1+\Bigg[\frac{(1-r_w^2-r_b^2-\Delta)(1-r_w^2-3r_b^2-\Delta)}{4r_b^2\, \Delta}\log\left(\frac{r_w^2}{1-r_w^2}\right)\Bigg]& \nonumber \\
&\qquad\qquad\qquad\qquad+\Bigg[\Delta\rightarrow -\Delta\Bigg]\Bigg\},&
\end{flalign}
\begin{equation}
I_0^{t\gamma w}= \Bigg[\frac{-1}{r_b\, \Delta}(1-r_b^2-r_w^2-\Delta)\log\left(\frac{1+r_w^2-r_b^2+\Delta}{2r_w}\right)\Bigg]+\Bigg[\Delta\rightarrow -\Delta\Bigg],\qquad\qquad\qquad\qquad\qquad
\end{equation}
\begin{flalign}
I_0^{bW\gamma}=
&\frac{2}{ r_b}+\Bigg\{\frac{ -2\left(1-r_b^2-r_w^2+\Delta \right)}{r_b\Delta(1+r_b^2-r_w^2+\Delta)^2}\;\Bigg[(1-r_w^2)^2-r_b^2(1+r_w^2)+\Delta(1-r_w^2)&\nonumber\\
&\qquad -2r_b^2(1-r_w^2+\Delta)\log\left(\frac{2r_b^2}{1-r_w^2-r_b^2+\Delta}\right)\Bigg]\Bigg\}+\Bigg\{\Delta\rightarrow -\Delta\Bigg\}&\nonumber\\
&\qquad +\Bigg[ 4\pi\, i\; r_b\; \frac{(1-r_w^2-r_b^2+\Delta)(1-r_w^2+\Delta)}{(1-r_w^2+r_b^2+\Delta)^2\Delta}\Bigg]-\Bigg[\Delta\rightarrow -\Delta\Bigg],&
\end{flalign}
\begin{flalign}
&I_0^{bw\gamma}= -2\; r_b \;\Bigg\{ \Bigg(\pi\,i \;\frac{1-r_b^2-r_w^2+\Delta}{\Delta(1+r_b^2-r_w^2+\Delta) }\Bigg) -\Bigg(\Delta\rightarrow -\Delta\Bigg)&
\nonumber\\ 
& \qquad+\Bigg[\frac{1-r_b^2-r_w^2+\Delta}{\Delta(1+r_b^2-r_w^2+\Delta)} \log\left(\frac{2 r_b^2}{1-r_b^2-r_w^2+\Delta}\right) \Bigg] +\Bigg[\Delta\rightarrow -\Delta\Bigg]\Bigg\},& 
\end{flalign}
\begin{equation}
I_0^{\gamma tb}=\frac{2\; r_b}{\Delta}\log\left(\frac{1-r_w^2+r_b^2+\Delta}{1-r_w^2+r_b^2-\Delta}\right),
\hspace*{.64\textwidth}\label{Igtb}
\end{equation}
with 
\[
\Delta=\sqrt{(1-r_w^2)^2+r_b^4-2r_b^2(1-r_w^2)}.
\]
In the limit $r_b \rightarrow 0$, the formulas get a simplest expression:
\begin{equation}
I_0^{t\gamma W}\xrightarrow{\ \ r_b\rightarrow 0\ \ } -\frac{r_b}{(1-r_w^2)^3}\left[(3-8r_w^2+5r_w^4)+4r_w^2(1-2r_w^2)\log(r_w)\right],
\end{equation}
\begin{equation}
I_0^{t\gamma w}\xrightarrow{\ \ r_b\rightarrow 0\ \ } \frac{2\; r_b}{(1-r_w^2)^2}\left[1-r_w^2+r_w^2\,\log(r_w^2)\right],
\end{equation}
\begin{flalign}
&\hspace*{1.cm}I_0^{bW\gamma} \xrightarrow{\ \ r_b\rightarrow 0\ \ } 2\;r_b\; \Bigg[\pi i\, \frac{\left(2+r_w^2+r_w^4\right)}{1-r_w^2} +
\frac{2\,r_w^2\left(1-2r_w^2+2r_w^6-r_w^8\right)}{(1-r_w^2)^4}\log(r_w)&\nonumber\\
&\qquad\qquad\qquad\qquad +1-\frac{2-r_w^2(5-4r_w^2+2r_w^2-2r_w+r_w^8)}{(1-r_w^2)^4}\log(1-r_w^2)\Bigg],&
\end{flalign}
\begin{equation}
I_0^{bw\gamma} \xrightarrow{\ \ r_b\rightarrow 0\ \ } \frac{-r_b}{1-r_w^2}\Bigg[2\pi\, i\,(1+r_w^2) +\log\left(\frac{r_b^2}{1-r_w^2}\right)
+r_w^2\log\left(\frac{r_w^2}{1-r_w^2}\right)\Bigg],
\end{equation}
\begin{equation}
I_0^{\gamma tb}\xrightarrow{\ \ r_b\rightarrow 0\ \ } -\frac{4\; r_b}{1-r_w^2}\log\left(\frac{r_b}{1-r_w^2}\right),\label{Igtb1}
\end{equation}
\end{appendices}
\bibliographystyle{JHEP}
\bibliography{BibLu_mod}

\end{document}